\documentclass[12pt]{article} 
\usepackage{amsmath} 
\usepackage{graphicx} 
 
\usepackage{natbib} 
\usepackage{psfrag} 
\usepackage{amssymb} 
 
\begin{document} 
 
\title{Deterministic and stochastic regimes of asexual evolution on rugged fitness landscapes}

\author{Kavita Jain$^{\S}$ and Joachim Krug$^{\dagger}$\\\mbox{}\\$^\S$Department of Physics of Complex Systems, \\The Weizmann Institute of Science, \\Rehovot 76100, Israel \\$^\dagger$Institut  
f\"ur Theoretische Physik,\\ Universit\"at zu K\"oln, 50937 K\"oln, Germany} 
 
\maketitle 
 
\newpage 
 
\noindent 
Running head: Dynamical regimes of asexual evolution 
\bigskip 
 
 
\bigskip 
 
\noindent 
Corresponding author: \\ 
Kavita Jain \\ 
Department of Physics of Complex Systems, \\ 
The Weizmann Institute of Science,\\ 
Rehovot 76100, Israel \\  
08-934-4983 (phone) \\  
08-934-4109 (fax) \\  
\texttt{kavita.jain@weizmann.ac.il} 
 
\bigskip

\noindent 
\textbf{Abstract:} We study the adaptation  
dynamics of an initially maladapted  
asexual population with genotypes represented by binary sequences  
of length $L$.  The population evolves in a  
maximally rugged fitness  
landscape with a large number of local optima. We find that whether the  
evolutionary trajectory is deterministic or stochastic depends on the  
effective mutational distance  
$d_{\mathrm{eff}}$ up to which the population can spread in genotype space.  
For $d_{\mathrm{eff}}=L$, the deterministic quasispecies theory operates  
while for $d_{\mathrm{eff}} < 1$, the evolution is completely stochastic.  
Between these two limiting cases, the dynamics are described by a local  
quasispecies theory below a crossover time $T_{\times}$ while  
above $T_{\times}$, the population gets trapped at  
a local fitness peak and manages to find a better peak either via stochastic 
tunneling or double mutations.   
In the stochastic regime $d_\mathrm{eff} < 1$, we identify two subregimes 
associated with clonal interference and uphill adaptive walks, respectively. 
We argue that our findings are relevant to the interpretation of  
evolution experiments with microbial populations.  
 
\newpage

The question whether the course of evolution is  
predetermined and if yes, to what  
extent and under what conditions this might be so has recently attracted  
attention of many researchers \citep{Wahl:2000, Rouzine:2001, Yedid:2002}.  
The answer to this question, particularly for large populations, is not  
obvious  
since the trajectories traced out during evolution 
are shaped by the interplay of the (deterministic) 
selective forces encoded in the fitness landscape, and  
the stochasticity of the mutational process, which limits  
the ability of the population to find and maintain favorable genotypes.

We address this question for an asexual population of size  
$N$ and binary genotype sequences of length $L$  
evolving on a fitness landscape.  
As there is a considerable evidence  
\citep{Whitlock:1995} for interactions amongst gene loci (or epistasis), it  
is important to consider the evolutionary process on a landscape  
that includes them. Such interactions  
may \citep{Wright:1932,Gavrilets:2004,Weinreich:2005b}  
or may not \citep{Lunzer:2005,Weinreich:2006} give  
rise to multiple peaks in the fitness landscape  
\citep{Jain:2006}. But at least on a qualitative  
level, recent experiments on microbial populations \citep{Elena:2003a}  
support the notion that 
the fitness landscape underlying the adaptive process has multiple peaks  
\citep{Lenski:1994,Korona:1994,Burch:1999,Burch:2000,Elena:2003b}.  
Motivated by this, we consider the  
dynamics of the evolutionary process on maximally rugged landscapes   
\citep{Kauffman:1987,Flyvbjerg:1992} which have high epistasis and a large  
number of adaptive peaks separated by valleys.

A detailed theoretical description of the evolution  
of a population subject to the combined 
effects of selection, mutation and stochastic drift in a complex  
fitness landscape constitutes a formidable problem, and previous 
studies have usually considered two limiting cases based on 
the size $N$ of the population and 
the mutation probability $\mu$ per generation per base (or gene locus). When 
the total number of mutants produced in a generation, $N L \mu$, is small,
the population consists  
of a single genotype at most times. Occasionally a mutation occurs in a  
single individual, 
which may become fixed in the population with a probability depending  
on the fitness advantage of the mutant. The population thus performs   
an \textit{adaptive 
walk} along a set of genotypes connected by single point mutations, 
which is biased towards high fitnesses and terminates at a local fitness  
maximum 
\citep{Gillespie:1984,Kauffman:1987,Macken:1989,Macken:1991,Flyvbjerg:1992, 
Orr:2002}. Clearly, the trajectory traced out by the population in this case  
is determined \emph {stochastically}.  
In the other extreme limit of $N \rightarrow \infty$ applicable to enormously  
large  
populations, each (relevant) genotype is populated by many 
individuals and the stochasticity inherent in the selection of individuals for 
reproduction can be neglected. This is the regime of \emph{deterministic}  
mutation-selection 
dynamics described by the quasispecies model, which was originally introduced  
in the 
context of prebiotic molecular evolution  
\citep{Eigen:1971,Eigen:1989,Baake:1999,Jain:2006}.   
 
Thus, in these two extreme cases either the population has many  
weighted paths  available or follows a single predetermined  
route to the global peak. One would like to know: what is the nature of the 
dynamics for parameters lying between these two limits?   
In the following section, we describe the model and introduce a parameter  
$d_{\mathrm{eff}}$ on the basis of which various dynamical regimes are  
distinguished. The effective distance $d_{\mathrm{eff}}$ is basically a  
measure of the extent to which a finite population can spread  
in the space of genotype sequences by mutations.  
For infinite populations, this distance equals  
the diameter $L$ of the entire sequence space, 
and we discuss this case in  
the section on quasispecies dynamics. 
We start with our earlier work on quasispecies evolution  
\citep{Krug:2003,Jain:2005} which provides in  
a suitably defined strong selection limit, a very transparent picture of the  
evolutionary trajectories and the genotypes that are encountered by a  
population moving towards the global fitness peak. We show that provided  
the mutation probability $\mu$ is sufficiently small, the analysis of  
\citet{Jain:2005} holds beyond the strong selection limit and the  
evolutionary trajectories obtained at different values of $\mu$ can be  
superimposed by a simple rescaling of time. The  
section on finite populations 
deals with the two subcases $1 \leq d_{\mathrm{eff}} < L$ and  
$d_{\mathrm{eff}} < 1$. The basic idea in the first case is that the finite  
population behaves like a quasispecies in an effective sequence space up to  
a certain timescale above which the stochastic evolution takes over. We  
estimate the time at which the crossover from deterministic to stochastic  
evolution occurs. For $d_{\mathrm{eff}} < 1$, the dynamics are  
stochastic at all times but depending on the product $N L \mu$, 
the dynamics may be characterized by the  
``clonal interference'' of several genotypes  
\citep{Gerrish:1998} or it may follow the 
adaptive walk scenario described above. 
In each case, we describe several individual fitness trajectories in detail  
both as a function of time and as a function of the system parameters.  
Finally, in the last section  
we summarize our results and discuss the relation of this work  
with that of others. 
 
\bigskip 
\centerline{MODELS} 
\bigskip 
 
We consider a haploid, asexual population with  
genotypes drawn from the space of binary sequences  
$\sigma = \{\sigma_1,...,\sigma_L \}$ of length $L$, where  
$\sigma_i = 0$ or 1. Depending on the context, a genotype can be thought  
to represent a small genome, a single gene or  
a sequence of $L$ biallelic genetic loci.  
A fitness $W(\sigma) \geq 0$ proportional to the expected number of offspring  
produced by an individual of genotype $\sigma$ is associated with 
each sequence. Reproduction occurs 
in discrete, non-overlapping generations. 
The structure of the population is monitored through the  
frequency $X(\sigma,t)$ of individuals of  
genotype $\sigma$ in generation $t$.

To simulate the stochastic evolution, a population of fixed size $N$ is 
propagated via standard Wright-Fisher sampling, i.e. each 
individual in the new population chooses an ancestor from the 
old population with a probability proportional to the fitness 
of the ancestor. Subsequently, point mutations are introduced 
with probability $\mu$ per locus per generation. In the  
limit of very large populations, this leads to a deterministic 
time evolution for the average frequency  
${\cal X}(\sigma,t) =  
\langle X (\sigma,t) \rangle$, where the angular brackets  
refer to an average over all realizations of the sampling process. 
The evolution equation reads as \citep{Jain:2006} 
\begin{equation} 
\label{quasi} 
{\cal X}(\sigma,t+1)= \frac{\sum_{\sigma'} p_{\sigma \leftarrow \sigma'}  
W(\sigma')  
{\cal X}(\sigma',t)}{\sum_{\sigma'} W(\sigma') {\cal X}(\sigma',t)}, 
\end{equation} 
where  
\begin{equation} 
\label{mut} 
p_{\sigma \leftarrow \sigma'} = \mu^{d(\sigma,\sigma')} (1 - \mu)^{L-d(\sigma,\sigma')} 
\end{equation} 
is 
the probability of producing $\sigma$ as a mutant of $\sigma'$ in one  
generation, and 
$d(\sigma,\sigma')$ denotes the Hamming distance between the two genotypes  
(i.e. the number 
of single point mutations in which they differ). Instead of simulating  
large (infinite) population, we numerically iterate  
the above discrete time equation.  
For future reference we note that the nonlinear evolution  
(\ref{quasi}) is equivalent to the linear iteration  
\begin{equation} 
\label{quasi_linear} 
{\cal Z}(\sigma,t+1)= \sum_{\sigma^{'}} p_{\sigma \leftarrow \sigma^{'}}  
W(\sigma')  
{\cal Z}(\sigma',t) 
\end{equation} 
for the unnormalized frequency ${\cal Z}(\sigma,t)$, where 
${\cal X}(\sigma,t) = {\cal Z}(\sigma,t)/\sum_{\sigma'} {\cal Z}(\sigma',t)$ 
\citep{Jain:2006}.

In order to generate a maximally rugged fitness landscape,  
the fitness values $W(\sigma)$ are chosen 
independently from a common  exponential distribution $P(W)=e^{-W}$  
with unit mean.  In the language of  
Kauffman's rugged landscape models  
in which the fitness contribution of each of the  
$L$ loci depends randomly on $K$ other loci, our uncorrelated landscape  
corresponds to $K=L-1$ and hence the limit of  
strong epistasis \citep{Kauffman:1993}. 
In particular, \textit{sign epistasis}, in the 
sense that a particular point mutation may be beneficial 
or deleterious depending on the genetic background 
\citep{Weinreich:2005b}, is common in these landscapes.  
We also note that the selection coefficient  
for a mutant of genotype $\sigma$ in a background of genotype 
$\sigma'$ is given by  
\begin{equation} 
\label{selection} 
s(\sigma,\sigma') = \frac{W(\sigma)}{W(\sigma')}-1, 
\end{equation} 
and the probability to find a genotype of fitness larger than $W$ is 
\begin{equation} 
\label{Q} 
Q(W) = \int_W^\infty dw \; P(w) = e^{-W}. 
\end{equation} 
 
We recall some typical properties of  
maximally rugged landscapes 
\citep{Kauffman:1987,Kauffman:1993}, which 
follow from elementary order statistics. For $S$ exponentially distributed  
random 
variables, the average value of the maximum is given by $\ln S$  
\citep{David:1970,Sornette:2000} which  
yields the expected fitness ${\cal W}_{\mathrm{max}} = L \ln 
2$ of the globally fittest among the $2^L$ genotypes. 
Correspondingly,  
the typical fitness of a local maximum which is a genotype  
without fitter one-mutant neighbors is  
${ \cal W}_{\mathrm{loc}} = \ln (L+1) \ll {\cal W}_{\mathrm{max}}$.  
Since the probability that a genotype is a local maximum  
is $1/(L+1)$, there are on an average $2^L/(L+1)$ local maxima in these  
landscapes. For such a genotype $\sigma$ with fitness  
${\cal W}_{\rm loc}$ surrounded by  
typical genotypes of fitness $W = {\cal{O}}(1)$,  
the selection coefficient $s(\sigma,\sigma') \sim  
\ln L \gg 1$. In this sense, we are dealing with  
a situation of \textit{strong selection} throughout this paper. 
 
For the purposes of illustration, we will base much of the 
discussion below on two  reference landscapes, each of which is a single  
realization of landscapes with sequence length $L = 15$ and $6$.  
The starting sequence $\sigma^{(0)}$ is a randomly chosen genotype 
at which the population finds itself in the beginning of the  
adaptation process. For our reference landscapes, $\sigma^{(0)}$ is  
of relatively poor fitness with value $W(\sigma^{(0)}) \approx 0.13$ 
for both cases. This has a rank $28795$ 
among $2^{15} = 32768$ genotypes and $55$ among $2^{6} = 64$ genotypes  
where the global maximum is assigned the rank $0$. The global peak  
is located at Hamming distance $10$ and $2$ from $\sigma^{(0)}$ with  
fitness  $W_{\mathrm{max}}=10.72$ and  
$4.29$ for $L=15$ and $L=6$, respectively.  
In the following discussion, instead of specifying actual fitness  
values for each sequence,  
we will provide their ranks as a subscript in the population density  
$X_{\mathrm{rank}} (\sigma,t)$.

In the subsequent sections, we will distinguish the dynamics on the 
basis of a parameter $d_{\mathrm{eff}}$ which is a measure of  
the typical extension of the population in genotype space  
and for strong selection, it is given by 
\begin{equation} 
\label{deff} 
d_{\mathrm {eff}} \approx \frac{\ln N}{\vert \ln \mu \vert} . 
\end{equation} 
Due to the 
quasispecies equation (\ref{quasi}), the average number of individuals 
produced in one generation at a sequence $\sigma$ located  
at distance 
$d(\sigma,\sigma^{(0)})$  from a localized population of size $N$   
is given by $N \mu^{d(\sigma,\sigma^{(0)})}$. Thus  
the maximum distance $d_{\mathrm {eff}}$ at which  
at least one individual (required for asexual reproduction)  
can be detected after one generation is given by (\ref{deff}).  
However in the next generation, the mutants of $\sigma^{(0)}$ can acquire  
further mutations thus extending the  
spread of the population beyond $d_{\rm{eff}}$.  We argue below that  
for landscapes with large selection coefficients as is the case  
with our rugged landscapes, the above definition is 
nevertheless a good approximation. 
 
To see this, let us consider the evolution in a landscape  
with infinite selection coefficients for which (\ref{deff}) is exact. As argued  
above, starting from a localised  
population at $\sigma^{(0)}$, at $t=1$ the population spreads over a typical  
distance  
$d_{\mathrm{eff}}$. If the landscape is such that all the sequences 
except the best one amongst the ones available within $d_{\mathrm{eff}}$ are lethal  
(i.e. with fitness zero), then in the next generation the  
population will move to the lone viable genotype  
(fitter than $\sigma^{(0)}$). This  
sequence in turn can be treated as the  
new $\sigma^{(0)}$ and the above argument can be applied recursively.  
 
That  
(\ref{deff}) cannot hold for weak selection can be seen by considering  
the flat fitness landscape (with selection coefficients zero) for which it  
is known that the average Hamming distance  
over which the population spreads is \citep{Derrida:1991} 
\begin{equation} 
d_{\mathrm {flat}} \approx \frac{L}{2} \left(\frac{4 \mu N}{1+4 \mu N} \right) , 
\end{equation} 
and which for large $N \mu$ is simply $L/2$. Away from these two limiting  
cases, one may expect an explicit dependence on the relevant selection  
coefficients. For rugged landscapes, one can get an idea of such a  
dependence at late times when  
(as explained in the section on finite populations),  
the population gets trapped at a sequence whose mutants within $d_{\rm{eff}}$  
are not fitter than itself. In such a case, the population at the peak  
and its surrounding valley reaches a stationary state and forms a  
quasispecies. Approximating the surrounding genotypes by a flat fitness  
landscape with $W(\sigma')=1$ and the localising sequence with fitness  
$W(\sigma) \gg 1$, an analysis within the unidirectional approximation shows that 
 the population distribution is an exponential \citep{Higgs:1994} 
\begin{equation} 
{\cal X}(\sigma') \sim \left(\frac{\mu (1+s(\sigma,\sigma'))}{s(\sigma,\sigma')}\right)^{d(\sigma,\sigma')} {\cal X}(\sigma). 
\end{equation} 
Defining $d_{\mathrm {eff}}$ as the genetic distance at which the  
population fraction falls to $1/N$, the resulting expression for  
$d_{\rm{eff}}$ is given by 
\begin{equation} 
d_{\rm{eff}} \approx \frac{\ln N}{|\ln \mu|} \left( 1+ \frac{1}{s |\ln \mu|} 
\right) 
\end{equation} 
which reduces to (\ref{deff}) with a correction that becomes 
negligible for $s(\sigma,\sigma') |\ln \mu| \gg 1$. When either the selection 
is weak or the mutation rate is large,  
the effective mutational distance is larger than  
given by (\ref{deff}).  
In the following sections, we will study three distinct cases classified on  
the basis of distance (\ref{deff}): (a) 
$d_{\mathrm {eff}}=L$ (b) $1 \leq  d_{\mathrm {eff}} <  L$ (c) $d_{\mathrm 
  {eff}} < 1$.  
 
\vspace*{1.cm} 
 
\bigskip 
\centerline{QUASISPECIES DYNAMICS} 
\bigskip

When the population $N \gtrsim \mu^{-L}$, the effective  
distance  $d_{\mathrm{eff}}=L$ and the population can spread all over the 
Hamming space. For small mutation probability $\mu$ (of the order 
$10^{-3}-10^{-8}$) that we consider here, this population size far exceeds the 
number of available genotypes. The requirement of such a 
large population size for a completely  
deterministic description comes from  
(\ref{mut}), according to which the mutation probability decreases  
exponentially with the distance.  
 
\begin{figure} 
\centerline{\includegraphics[width=14cm,angle=270]{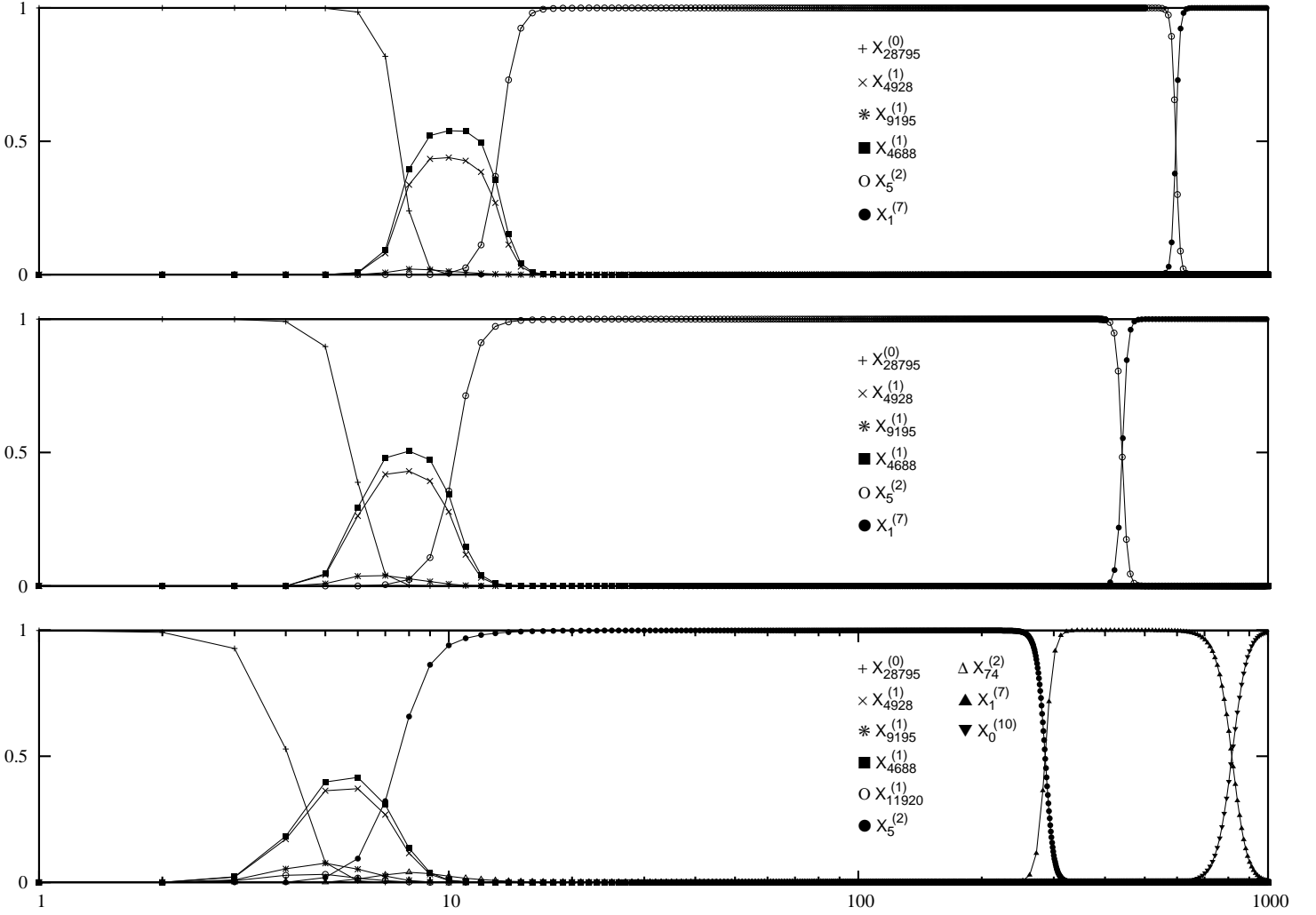}} 
\caption{\label{fig:QN15}Quasispecies evolution of the population  
${\cal X}_{\mathrm {rank}}^{d(\sigma,\sigma^{(0)})}$. The numerical iteration  
of equation (\ref{quasi}) is shown for $\mu=10^{-8}$, $10^{-6}, 
  10^{-4}$ (top to bottom) with $L=15$, starting from  
all the population at sequence $\sigma^{(0)}$ in the fitness 
landscape explained in the text. The sequences with fraction  
$\geq 0.005$ are shown.} 
\end{figure}

The discrete time quasispecies equation (\ref{quasi}) 
was iterated numerically for the population fraction  
${\cal X}_{\mathrm{rank}}^{d(\sigma,\sigma^{(0)})}$, where we have labeled a  
sequence  
by its rank and Hamming distance from $\sigma^{(0)}$. The  
time evolution is depicted in Figure \ref{fig:QN15}  
for various $\mu$ and fixed $L$.  
Since ${\cal X}(\sigma,1) \sim \mu^{d(\sigma,\sigma^{(0)})}$ , all 
the mutants become available immediately with a concentration  
decreasing exponentially with the distance from the parent 
sequence. As a result, the population at fitter sequences closer to  
the parent increases and that at $\sigma^{(0)}$ 
decreases. One of these fit sequences becomes dominant in the sense  
that it supports the largest population. This sequence is in turn overtaken  
by a fitter sequence close to it, and  
this process of leadership changes goes on until the population has reached the 
global maximum. We are interested in the evolutionary trajectory traced out by  
the most populated sequence $\sigma^{*}(t)$ at time $t$.  
 
The analysis of \citet{Krug:2003} and \citet{Jain:2005} provides a simple  
way of  
identifying the genotype $\sigma^{*}$ for a given landscape and a  
given starting sequence.  
It is based on a particular strong selection limit, in which 
the mutation rate is scaled to zero and the fitnesses are scaled 
to infinity in such a way that the (appropriately normalized)  
logarithmic population fractions remain well behaved. The key  
observation is that the behavior of the evolutionary trajectory 
$\sigma^\ast(t)$ can be accurately predicted by simply assuming that  
the mutations can be turned off once the sequence space has been  
``seeded'' by the population fraction $\sim \mu^d$ that is established  
by mutations after the first generation. Thus, 
each unnormalised population frequency ${\cal Z}(\sigma, t)$ changes  
exponentially in time according  
to its own fitness,  
from an initial value proportional to  
$\mu^{d(\sigma, \sigma^{(0)})}$. 
In logarithmic variables, this implies the simple 
linear time evolution (see also \citet{Zhang:1997})  
\begin{equation} 
\label{lines} 
\ln {\cal Z}(\sigma,t)= - \vert \ln \mu \vert d(\sigma, \sigma^{(0)})+  
\ln W(\sigma) \; t . 
\end{equation} 
Since the first term on the right hand side  
is the same for all sequences in a \textit{shell} of constant 
Hamming distance $d(\sigma,\sigma^{(0)})$, within each shell only the sequence 
with the largest fitness needs to be considered for  determining  
$\sigma^\ast(t)$. It is also evident from  (\ref{lines}) that 
among these shell fitnesses only fitness \textit{records}, that is, sequences 
whose fitness is larger than the fitnesses in all shells closer to  
$\sigma^{(0)}$, 
can possibly partake in the evolutionary trajectory. Fitness records can be  
identified purely on the basis of the fitness rank. Their statistical  
properties 
are independent of the underlying fitness distribution, but depend on the  
geometry 
of the sequence space \citep{Jain:2005,Krug:2005}.  
 
The set of sequences $\{ \sigma^{*}\}$ making up the evolutionary trajectory  
is  
a subset of the fitness records, from which those records are eliminated 
which are being \textit{bypassed} by a fitter but more distant record before 
reaching the status of the most populated genotype \citep{Krug:2003,Sire:2006}.  
To decide whether a given 
record is bypassed, the actual fitness values and not just their ranks 
are needed. Bypassing is a significant effect: it reduces the number of 
steps in the evolutionary trajectory from the number of records, 
which is of order $L$, to the order 
$\sqrt{L}$ for logarithmic fitness distributions of the exponential  
type \citep{Jain:2005,Krug:2005}. Thus, not all of the $L+1$   
mutant classes can appear  
in the trajectory and in fact, only a vanishing fraction of a total  
of $2^L$ genotypes actually appear (Figure \ref{fig:QN15}).

\begin{figure} 
\psfrag{ylabel}{${\cal W}(t)$} 
\centerline{\includegraphics[width=8cm,angle=270]{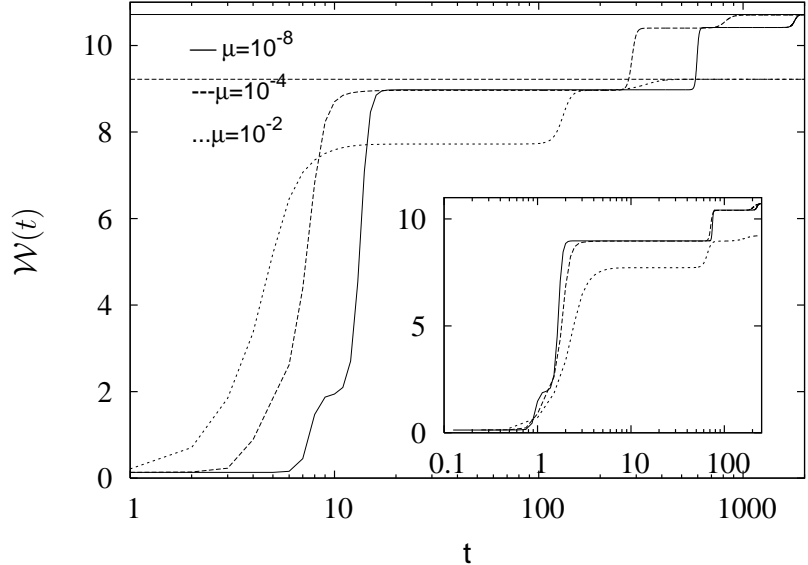}} 
\caption{\label{fig:avgQ}Punctuated rise of the average fitness ${\cal W}(t)$ 
for fixed landscape and fixed initial condition in the  
quasispecies model with genome length $L=15$. The solid line is the  
  fitness $W_{\mathrm{max}}$ of the global maximum and broken one is  
$e^{-\mu L} W_{\mathrm{max}}$ with $\mu=10^{-2}$. The steps become more  
diffuse as $\mu$ increases, and the fitness level is reduced for the largest 
value of $\mu$ due to the broadening of the genotype distribution. 
Inset: Average fitness plotted 
  as a function of $t/|\ln \mu|$ to show the scaling of jump times. } 
\label{avgQ} 
\end{figure} 
 
When applied to our reference landscape, the above analysis predicts an 
evolutionary trajectory involving the genotypes with ranks $28795$, $4688$,  
$5$, $1$ and $0$ in shells $0, 1, 2, 7$ and $10$ respectively,  
which are precisely the ones that appear in Figure \ref{fig:QN15}. Each of  
these genotypes is also a record, none of which is bypassed in the landscape  
used here. To see bypassing of the contending genotypes, we need to consider  
larger $L$ as the number of bypassed sequences increases with $L$.  
 
As Figure \ref{fig:QN15} shows, although  
the set $\{ \sigma^{*} \}$ remains the same  
for a broad range of mutation probability $\mu$,  
the timing of the appearance of new mutants and the polymorphism  
of the population depends on $\mu$. These effects are also  
reflected in the stepwise behavior of the  
population averaged fitness  
${\cal W}(t)=\sum_{\sigma} W(\sigma) {\cal X}(\sigma, t)$  
in Figure \ref{fig:avgQ}. For smaller $\mu$, adaptive events  
occur at later times.  
This is expected on the 
basis of (\ref{lines}) from which $\mu$ can be eliminated by a  
rescaling of time with $\vert \ln \mu \vert$. Indeed, the inset  
shows that the 
timing of the peak shifts can be made 
to coincide by scaling time with $\vert \ln \mu \vert$. The other  
effect with increasing $\mu$ is that the  
transitions between fitness peaks become 
more gradual \citep{Krug:1993}, and the fitness level at a  
given (rescaled) time is lowered. This happens due to an increase in the  
diversity (the number of genotypes present in the population)   
which  
is controlled by the probability $1-(1-\mu)^L \approx \mu L$ for any mutation  
to occur. For the largest mutation probability $\mu=10^{-2}$ that we consider  
here, this probability is significant and the mutational load  
\citep{Haldane:1927} can be estimated as follows. Using the quasispecies equation (\ref{quasi}) in the steady state within unidirectional approximation \citep{Jain:2006} for the master sequence with fitness  
$W(\sigma^\ast)$, it immediately follows that the population  fitness is 
given by $W(\sigma^\ast) e^{-\mu L}$ for large $L$ and small $\mu$.  
The mutational load is thus $W(\sigma^{*}) \; (1-e^{-\mu L})$ and the fitness  
is reduced by a factor $e^{-\mu L} \approx 0.86$ 
for $\mu = 0.01$ and $L=15$, in very good agreement with the data in 
Figure \ref{fig:avgQ}. To summarise, the mutations affect the dynamics  
in two respects: on decreasing $\mu$, the new mutants get fitter but are  
slower to appear (``slow-but-fit'').

\bigskip 
\centerline{FINITE POPULATIONS} 
\bigskip 
 
As we discussed above, in the infinite population limit  
all the genotypes are immediately occupied so that the subsequent   
dynamics involving the fit genotypes can be approximated as largely due to the  
selection process.  
For finite $N$ on the other hand, the  
population distribution has a finite support $d_{\mathrm {eff}}$ at 
any time. Then if the distance to the genotype that offers 
selective advantage 
over the currently dominant one is larger than $d_{\mathrm {eff}}$, or 
the distance $d_{\mathrm {eff}}$ is less than unity,  
the average number of individuals at the desired 
distance is smaller than one.  
One cannot work with averages under such circumstances and must take 
fluctuations arising due to rare mutations into account.  
 
\begin{figure} 
\centerline{\includegraphics[width=14cm,angle=270]{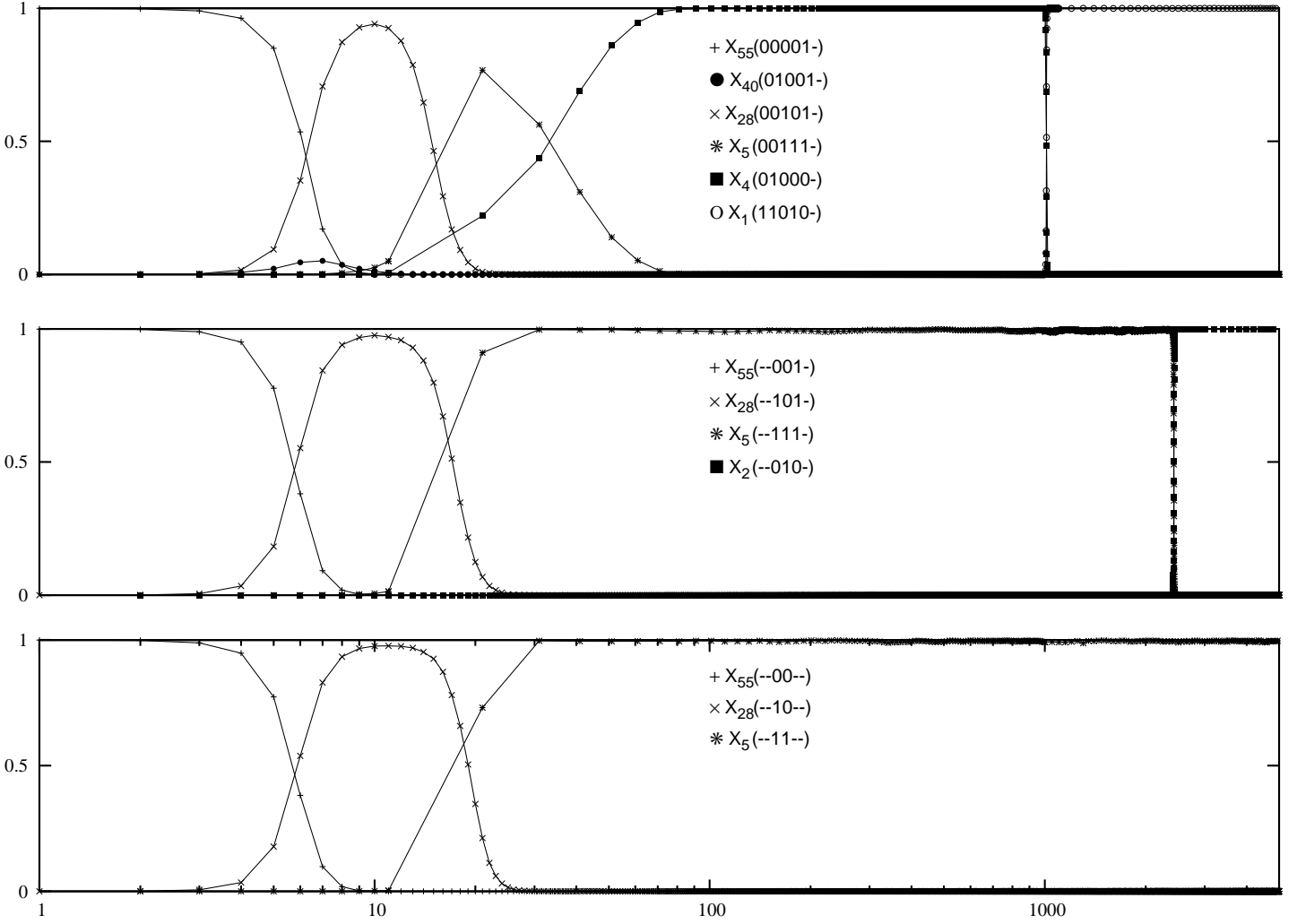}} 
\caption{\label{fig:M14m4}Evolutionary trajectories in a sequence space of  
length $L=6$ 
 with $N=2^{14}, \mu=10^{-4}$ so that $N \mu \approx 1.64$ and  
$d_{\mathrm{eff}} \approx 1.05$. The 
 population fraction is denoted by 
 ${X}_{\mathrm{rank}}(\sigma)$ where the $\sigma_{i}$'s that do not 
 change in the course of time are represented by a dash.  
Only the sequences with population fraction $\geq 0.05$ are  
shown. In the initial phase, the  
three populations ${X}_{55}, {X}_{28}$ and ${X}_5$ occur in all of the 
 above trajectories and have rather similar curves supporting  
deterministic evolution. At late times, the population escapes the local peak  
with rank $5$  
via tunneling in the top panel and by a double mutation  
in the middle one.} 
\end{figure} 
 
\begin{figure} 
\centerline{\includegraphics[width=5.8cm,angle=270]{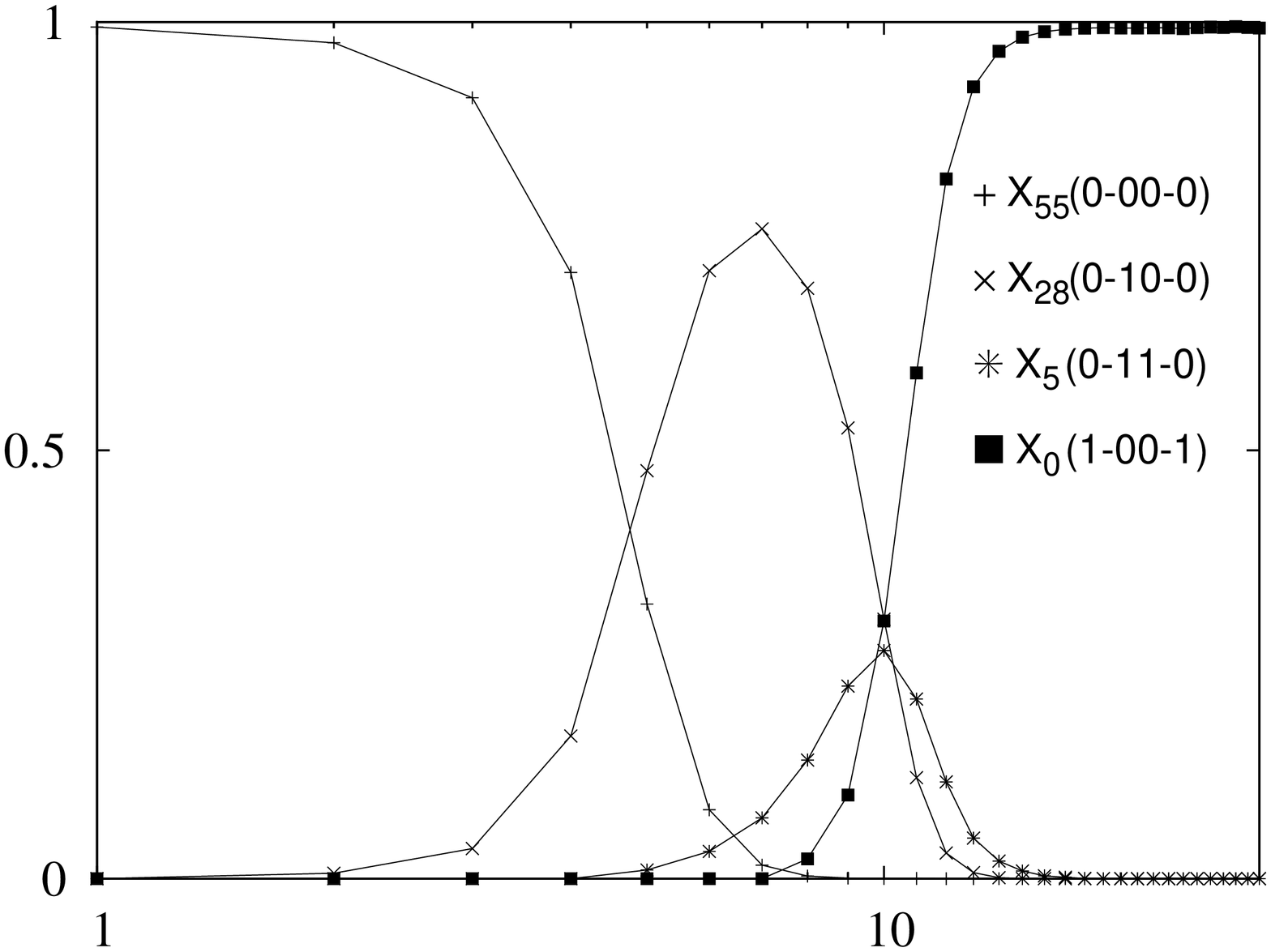} 
\includegraphics[width=5.8cm,angle=270]{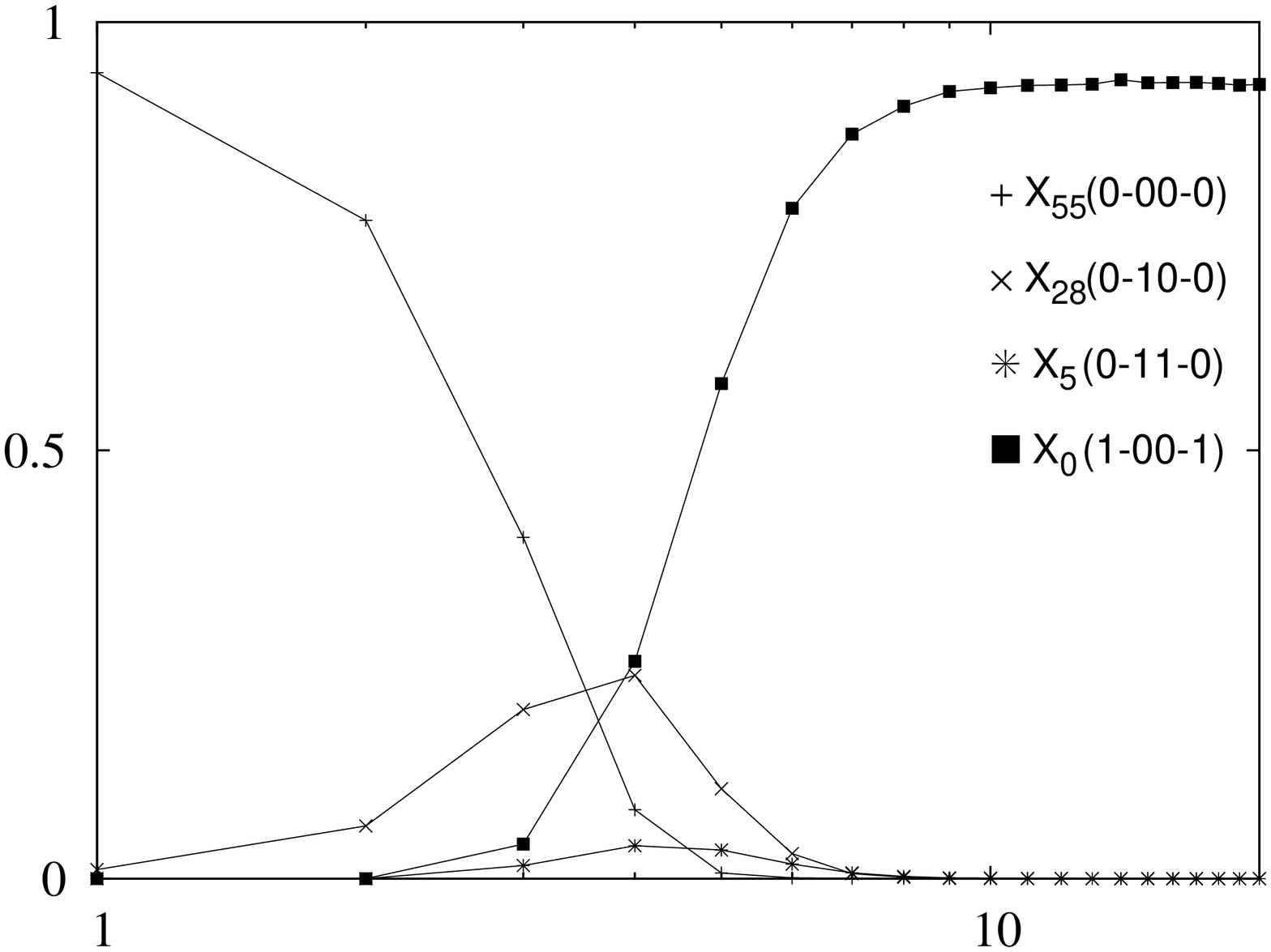}} 
\caption{\label{fig:M14} Evolutionary trajectories for $\mu=10^{-3}$ (left)  
and $\mu=10^{-2}$ (right) with $N=2^{14}$ and $L=6$. In the left panel,  
the effective distance $d_{\rm{eff}} \approx 1.4$  
and the population  
passes deterministically through the rank 28 sequence 
towards the global maximum.   
In the right panel, $d_{\rm{eff}} \approx 2.1$ and the  
population reaches the global maximum almost immediately.} 
\end{figure}

{\bf Crossover from deterministic to stochastic dynamics:}  
We first consider the case when $1 \leq d_{\mathrm{eff}} < L$. Starting from a  
parent sequence $\sigma^{(0)}$ supporting a population $N \ll 
\mu^{-L}$, the mutants can spread up to a shell at a distance 
$d_{\mathrm {eff}} < L$. Then provided the selection coefficient involving the  
fittest and the next few fittest sequences within $d_{\rm{eff}}$ is large,  
 the dynamics within this distance are similar to  
the 
quasispecies case in that the population at the fittest sequence in each shell  
competes with 
the one in other occupied shells, and passing through sequences at 
which it becomes dominant in the least time finds the best 
available sequence $\sigma^{*}$ within $d_{\mathrm {eff}}$ of $\sigma^{(0)}$.  
The last step is akin to finding the global maximum in the quasispecies 
case. If however, the selection is not strong, several fit genotypes get 
populated, and due to a mutation in this  
set of fit sequences, the population may be able to find a sequence even  
fitter than the fittest sequence $\sigma^*$ within $d_{\rm{eff}}$. In such an  
event, the fittest sequence within  
$d_{\rm{eff}}$ still achieves a majority status but only momentarily.  
Similar process is repeated within shells at radius $d_{\mathrm 
  {eff}}$ from the new most populated sequence $\sigma^{*}$. The above  
 deterministic 
process is expected to occur for individual  
trajectories obtained in stochastic simulations as long as the 
population can find a sequence better than the current $\sigma^{*}$  
within a distance $d_{\mathrm{eff}}$. In 
particular, for $d_{\mathrm{eff}} \sim 1$, the \emph{local quasispecies}  
evolution continues until the population hits a local peak, after which  
stochastic evolution takes 
over. The latter typically involves ``crossing the valley'' via 
less fit nearest neighbor mutants to a better peak than the current one.  
 
In Figure \ref{fig:M14m4}, we chose $d_{\mathrm{eff}}$ slightly above 
unity; since at any time, typically the population can sense only $L$ 
sequences, we work with a small sequence space of length $L=6$ to 
reduce the number $2^L-L$ of unoccupied sequences.  
Also, we keep the mutation probability $\mu$ somewhat large  
since for $d_{\mathrm{eff}}$ close to one,   
$N \sim \mu^{-1}$ and Wright-Fisher sampling requires operations 
of order $N$ per time step.  
Note that in this case the number of genotypes $2^L = 64$ is much smaller than 
the population size. Nevertheless, we will see that the dynamics is far from  
the 
deterministic quasispecies limit, because the more stringent condition 
$d_\mathrm{eff} = L$ is not met. Since doubling $d_\mathrm{eff}$ requires increasing 
the population size from $N$ to $N^2$, it is clear that fully deterministic 
behavior can be realized only under extreme conditions.  
 
{\noindent \it Deterministic dynamics:}  
The different runs in Figure \ref{fig:M14m4} correspond to 
different sampling noise with all the other parameters  
kept the same. We  
start with all the individuals at sequence $\sigma^{(0)}$ with rank 
$55$. Since $d_{\mathrm{eff}}$ is close to one, the population spreads 
from here to sequences within Hamming distance unity of $\sigma^{(0)}$ 
and moves to the best sequence amongst them namely the sequence with rank 
$28$. In this case, there is no bypassing (discussed in  
the quasispecies section) of a fit sequence  
and the best sequence in the first shell becomes the most populated  
sequence $\sigma^*$. As the population  
at this sequence grows, the chance that it will produce its  
one-mutant neighbors also increases; in fact, a  
mutant $\tilde \sigma$  
better than $\sigma^*$  
appears at time $\tau \sim (1/s) \ln (s/N \mu^2)$ where the selection  
coefficient $s=s(\tilde \sigma,\sigma^\ast$),  
when the fraction at the current $\sigma^*$ becomes $\sim 1/N \mu$  
\citep{Wahl:2000}. The population then starts growing at  
the sequence with rank $5$ which is the  
best sequence in the first shell centred about the sequence ranked 
$28$. The process so far is deterministic as is evident from the three   
runs. Note that the set $\sigma^{*}$ obtained using the local  
quasispecies theory will in general be different from the  
quasispecies analysis of the Hamming space  
containing all shells up to the shell  
in which the local peak is situated; this is because  
the sequences obtained in the former case 
can be outcompeted by fitter mutants  
before reaching fixation as discussed in the last section. For instance, if we  
apply the  
deterministic prescription to the Hamming space restricted to shell  
$2$ about $\sigma^{(0)}$, the sequence ranked $5$ will not appear in the  
trajectory since it  
will be immediately overtaken by the global maximum which also lies  
in shell $2$.  
 
In Figure \ref{fig:M14m4}, the sequence with rank $5$ is a local peak so 
a better sequence lies beyond distance unity; in fact, it lies in 
the second shell about this local peak and carries the rank label $2$.  
The trajectories in Figure \ref{fig:M14m4} take different routes 
from here onwards. In all the three  
cases, the last most populated sequence shown is at a distance $4$ from the  
global maximum, which in fact lies at distance $2$ from the initial sequence.  
Thus, a finite population wanders around and is inefficient  
in search of the global peak.  
 
Figure \ref{fig:M14} shows the evolutionary trajectories for larger  
$\mu$ (and hence $d_{\rm{eff}}$) for fixed population size. In the left  
panel, since $d_{\rm{eff}} \approx 1.4$, the population finds the best sequence  
$28$ in shell one about $\sigma^{(0)}$ as before. But as the sequence with  
globally largest fitness  
became available due to a mutation in a  
nearest neighbor mutant of $\sigma^{(0)}$, the population moves to the global  
peak. We performed several runs for this set of parameters  
and found that $X_5$ never achieved a majority  
status. On increasing $\mu$ further corresponding to  
$d_{\rm{eff}} \approx  2.1$, the sequence with rank $0$ being within  
$d_{\rm{eff}}$ of the initial sequence became immediately available, and the  
population formed a quasispecies around the global peak.

{\noindent \it Stochastic dynamics:}  
We now describe the individual trajectories  
in Figure \ref{fig:M14m4} in some detail. In the top panel, at $t=7$,  
a nearest neighbor of $\sigma^{(0)}$ with rank $40$ mutated at one locus to  
produce an individual at rank $4$ sequence which is a local peak.  
The rank 4 sequence replaces the rank 5 sequence as the most populated  
genotype before the rank 5 sequence has reached fixation. Since the two 
sequences are 4 point mutations apart, this constitutes an example 
of what has been called a \textit{leapfrog} episode, in which two 
consecutive majority genotypes appear that are not closely related 
to each other but have a common ancestor further back in the 
genealogy \citep{Gerrish:1998}.   
Later, a rank $50$ neighbor of rank $4$ sequence mutated once at $t=996$  
to populate rank $1$ sequence thus enabling the population to shift from  
one peak to another.  
 
In the middle panel, although a rank $48$  
neighbor of the sequence ranked $5$ mutated once at $t=1234$ to produce an  
offspring with rank $2$, this individual was lost. At $t=2384$, a double  
mutation in the  
sequence ranked $5$ allowed the population to shift the peaks without  
crossing the valley. In the last panel, the population remained trapped  
at the rank $5$ sequence until the last observed time $t=10^4$.  
 
The process  
of shifting peaks via valley crossing \citep{Wright:1932} or stochastic  
tunneling \citep{Iwasa:2004,Weinreich:2005} can happen if  
many mutants at Hamming distance unity from a local peak are available.   
While the Wrightian concept of valley crossing involves moving  
the whole population through a low fitness sequence, the process of  
stochastic tunneling only requires the  
presence of a few low fitness neighbors and we discuss this here.  
During the residence time of the population near the peak,  
a mutation-selection balance is reached  
between the peak genotype and its one-mutant neighbors.  
Then the average fraction of population at a given valley sequence with  
fitness $W_{\mathrm{mut}}$  
can be estimated using the quasispecies equation, and one has  
\begin{equation} 
\label{mutsel} 
{\cal{X}}_{\mathrm{mut}} \approx \frac{\mu W_{\mathrm{mut}}}{W_{\mathrm{loc}}-W_{\mathrm{mut}}}  
\end{equation} 
where $W_{\mathrm{loc}}$ is the fitness of the local peak.  
Clearly, the total number of  mutants produced  
depends on the neighborhood of the local peak; if the fitness of the  
neighbors is much smaller  
than that of the local peak, then it is of the order  
$N \mu L/W_{\mathrm{loc}}$ on using  
that the average value of exponentially distributed variables is $1$.  
Else it is dominated by the population at the best one-mutant neighbor with  
fitness close to $W_{\mathrm{loc}}$.  
In Figure \ref{fig:M14m4}, the sequence ranked $4$ produced  
on average $N L \mu \sim 10$ mutants, while rank $5$ produced a suite of  
about $200$ mutants, a lower bound ($\sim 80$) on which can be obtained by  
using  
(\ref{mutsel}) and the fitness $W_{\mathrm{mut}}$ of the rank $6$ sequence,   
which is the fittest nearest neighbor of rank $5$ sequence.  
 
Since there are typically many low fitness sequences available  
in the valley, it is likely  
that the population trapped at a local peak escapes due to a mutation  
in one of the $N \mu L$ one-mutant neighbors. This gives the  
simple estimate of the tunneling time to be $\sim (N \mu^2 L)^{-1} \sim 10^3$  
for our choice of parameters. This in fact is a lower bound as the  
tunneling time depends inversely on the advantage conferred by the  
next local peak.  
An expression for the rate ($\sim T_{\mathrm{tunnel}}^{-1}$)  
to tunnel to a beneficial mutation via a  
deleterious one has been obtained in \citet{Iwasa:2004} using a Moran  
process (also see \citet{Weinreich:2005}). This is given  
by the product of three factors:  
average number of deleterious mutants produced, mutation  
probability with which a deleterious mutates to an advantageous one and the  
fixation probability which is  
the relative fitness difference between the final and initial mutants finally   
yielding   
\begin{equation} 
\label{Ttunnel} 
T_{\mathrm{tunnel}} \sim \left [N \mu^2 L  
\left(\frac{1}{W_{\mathrm{loc,i}}}-\frac{1}{W_{\mathrm{loc,f}}} \right)  
\right]^{-1} 
\end{equation} 
where $W_{\mathrm{loc,\{i,f \}}}$ refers to fitness of the  
initial and the final local peaks.  
Inserting the fitness values of the two local peaks in question,  
$T_{\mathrm{tunnel}}$ turns out to be $\sim  3000$ which is somewhat  
larger than that observed in the top panel.  
 
In the middle panel, although  
many mutants are available at the valley sequence ranked $6$, the  
population could not tunnel through this sequence as it does not  
have a better neighbor other than sequence $5$ itself.  
Instead a double mutation at $t=757$  
was responsible for escaping the local  
peak at sequence ranked $5$ to the next local peak with rank $2$.  
Since the time $T_{\mathrm{double}}$ for the (desired) double mutation  
to occur in one generation is given by \citep{Iwasa:2004,Weinreich:2005} 
\begin{equation} 
T_{\mathrm{double}} \sim \frac{W_{\mathrm{loc}}}{W_{\mathrm{loc}}-W_{\mathrm{mut}}} T_{\mathrm{tunnel}}, 
\end{equation} 
it exceeds $T_{\mathrm{tunnel}}$  
if $W_{\mathrm{mut}} \sim W_{\mathrm{loc}}$,  
and in such a case, tunneling is the dominant mode of escaping the local peak.  
On the other hand, the valleys typically encountered in a rugged landscape  
are ``deep'' as ${\cal W}_{\mathrm{mut}}=1$ and  
${\cal W}_{\mathrm{loc}}=\ln L$. In this situation, the population may attempt  
to hop across the valley; the  
probability for such an event is roughly given by $N \mu^2$ times  
the average number of fitter neighbors available at distance $2$ away.  
The latter is simply $(L^2/2)\; Q(W(\sigma^{*}))$.  
Using  
${\cal{W_{\mathrm{loc}}}}=\ln L$, we again find that the time scale over  
which a double mutation can occur is of the same order as the tunneling time. 
 
{\noindent \it Crossover time:}  
We now estimate the time $T_{\times}$  
at which the crossover from deterministic to  
stochastic evolution occurs using an argument employed  
previously by \citet{Krug:2003} and \citet{Jain:2005}.  
We consider the evolution  
equation (\ref{lines}) for the unnormalised population  
according to which the logarithmic population at a fit sequence increases  
linearly. Then the crossover time $T_{\times}$ at which  
the first local peak is reached can be approximated by the typical time at  
which the population at the first  
local peak (rank $5$ in Figure \ref{fig:M14m4}) overtakes the population  
at the most populated sequence  
$\sigma^{*}$ (rank $28$) at Hamming distance unity from it. 
This is given by 
\begin{equation} 
T_{\times} \sim \frac{|\ln \mu|}{\ln (W_{\mathrm{loc}}/W(\sigma^{*}))}.  
\end{equation} 
For the landscape  
used in Figure \ref{fig:M14m4}, the fitness $W(\sigma^*) \approx 0.81$  
and $W_{\mathrm{loc}} \approx 1.65$ so that $T_{\times}$ works out to be  
about $13$ time steps which is in reasonable agreement with the  
time at which $X_{5}$ appears.  
The dependence of $T_{\times}$ on $L$ can  
be found by  
noting that generally the fitness ratio in the argument of the logarithm 
is close to unity, so that the logarithm can be expanded. 
The denominator then  
reduces to $W_\mathrm{loc}/W(\sigma^\ast) - 1 \approx 1/W(\sigma^{*})$  
on using that the typical difference between two exponentially distributed 
independent random variables is equal to unity 
 \citep{David:1970,Sornette:2000}. The fitness $W(\sigma^{*})$ of the  
last-but-one most populated sequence  
in the quasispecies regime is expected to be of the same  
order as the fitness of the local peak which  
increases as $\ln L$.  
Thus for exponentially distributed fitness and $d_{\mathrm{eff}}=1$,  
the local quasispecies theory works over a time scale that increases as 
\begin{equation} 
\label{Ttimes} 
T_{\times} \sim |\ln \mu| \; \ln L . 
\end{equation}

Although we mainly discussed the case $d_{\mathrm{eff}}=1$ above, it is  
easy to see that for larger effective distance also, the local  
quasispecies theory will work up to a crossover time after which the  
population will get trapped at a ``local peak''  
which does not have a better  
sequence available within Hamming distance $d_{\mathrm{eff}}$ and will have  
to wait for a rare mutation to find a better sequence.  
For $d_{\mathrm{eff}} \ll L$, the crossover time  
can be easily generalised by approximating it by the time required for  
the last overtaking event to happen which is given by 
\citep{Krug:2003,Jain:2005} 
\begin{equation} 
T_{\times}(d_{\mathrm{eff}}) \sim \frac{|\ln \mu| d_{\mathrm{eff}}} 
{\ln (W_\mathrm{loc}/W(\sigma^\ast))}. 
\end{equation} 
Expanding the logarithm as above, and  
using that the peak genotype  
is the best amongst $\sim L^{d_{\mathrm{eff}}}$ sequences, it  
follows that 
\begin{equation} 
\label{Ttimesf} 
T_{\times}(d_{\mathrm{eff}}) \sim d_{\mathrm{eff}}^2 \; |\ln \mu| \; \ln L.  
\end{equation}

{\bf Fully stochastic evolution:}  
We now turn to the regime when the effective  
distance is less than unity. Unlike in the previous cases, now the dynamics  
is stochastic at all times.  
The parameter $d_{\mathrm {eff}} < 1$ 
implies that the average number of mutants $N \mu$ produced at Hamming  
distance unity is also smaller than $1$. Since the population is  
discrete, this number cannot be observed until time $\sim (N 
\mu)^{-1}$ when one mutant is produced at a given 
sequence. However, since the mutation probability is rotationally 
symmetric, a total of $\sim L N \mu$ new mutants at Hamming distance 
unity can be produced in one generation. The dynamics depend on whether  
the parameter $L N \mu$ is above or below unity, and we study these  
two cases in the following subsections. We will mainly focus on the  
short time regime as the behavior at long times is expected to be similar  
to that discussed previously. 
 
\begin{figure} 
\centerline{\includegraphics[width=14cm,angle=270]{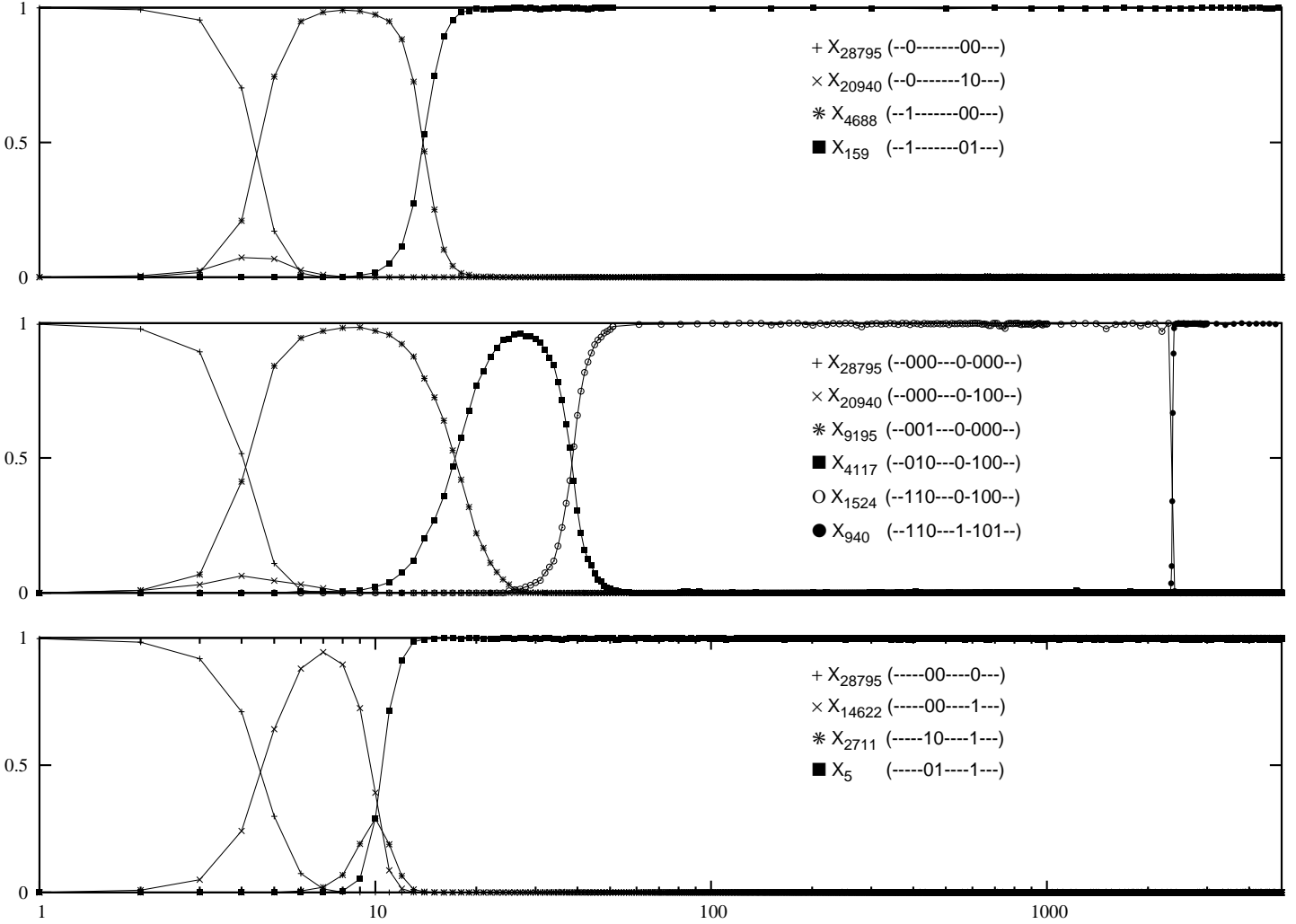}} 
\caption{\label{fig:M10m4N15}Stochastic trajectories for  
$L=15, N=2^{10}, \mu=10^{-4}$ with  
$N \mu \approx 0.10$ and $L N \mu \approx 1.54$. The population 
passes through different routes in each case right from the beginning 
and at short times, several mutants at constant Hamming distance are  
produced simultaneously. Only the mutants that achieve a  
fraction $\geq 0.005$ are shown in the plot. In the top panel, all the mutants  
shown belong to the same lineage; in the next two panels, while a fit  
mutant is on its way to fixation, a split in the lineage produced even better  
mutant thus bypassing the former one.} 
\end{figure}

{\noindent \it Clonal interference:} Figure \ref{fig:M10m4N15} shows the  
temporal evolution  
of the population fraction for three different sampling noise (keeping  
rest of the parameters same). Clearly the population traces different 
trajectories in each case. In Figure \ref{fig:M10m4N15}, the population at  
$\sigma^{(0)}$ produced a total of  
$N L \mu \sim 1-2$ mutants in one generation. Thus in this regime, the  
sequence space is very sparsely populated as only $2$ to $3$ genotypes  
are occupied. But since many (about $L Q(W(\sigma^{(0)})) \sim 13$) of them  
are  
better than the parent, the population immediately begins the hill-climbing  
process. In the top panel of Figure \ref{fig:M10m4N15}, the best  
one-mutant neighbor of $\sigma^{(0)}$ with rank $4688$ mutated once at  
$t=6$ to move the population at a highly fit sequence ranked $159$ which is  
also a local peak. In the middle panel, while most of the population climbed  
the nearest neighbor of parent with rank $9195$, an individual at a much lower  
rank $20940$ produced an offspring at $4117$ at $t=5$. Thus, due to  
the interference of rank $20940$ sequence, the population managed to access  
an even fitter sequence. After a single mutation at the genotype ranked  
$4117$, the population reached a local peak with rank $1524$ from where  
it escaped via double mutation. In the last panel, at $t=5$, the rank $14622$  
neighbor of $\sigma^{(0)}$ mutated once to populate a local  
peak with rank $2711$. However, the population escaped this local peak  
by {\it climbing} a better local peak with rank $5$ made available due to  
one mutation in sequence $14622$ at $t=7$. In each case, since the selection  
coefficients involved are of order unity, the fitter mutants get fixed immediately and one can neglect  
the time to reach fixation.  
 
In the preceding sections with $d_{\mathrm{eff}} \gtrsim 1$, all the  
mutants are available within the occupied shells and the best amongst them  
becomes the most populated sequence $\sigma^*$. However, for $N \mu < 1$,  
only a few randomly sampled sequences can get populated and as most  
of the genotypes available at Hamming distance one from $\sigma^{(0)}$ 
are of comparable fitness,  
each of them can achieve a moderate population frequency.  
While the best amongst them  
has the highest chance of achieving majority status,  
the other mutants in the meanwhile can establish their own lineage by creating  
their own (small) suite of one-mutant neighbors. If a mutant better than  
the one that is currently going to fixation is produced, there is a  
competition and the latter is bypassed.  
This process is reminiscent of the bypassing discussed in  
the quasispecies section  
- in both the cases, while a fit mutant is going to fixation, it  
may get bypassed by an even better one. However, while the set of  
mutants that will compete with each other in this manner is predetermined  
for large populations, here they are stochastically generated in time.

The competition between several beneficial mutations in an asexual population 
has been termed \textit{clonal interference} \citep{Gerrish:1998}.  
A quantitative criterion for the occurrence of clonal interference, 
adapted to the present situation, reads \citep{Wilke:2004}  
\begin{equation} 
\label{clonal} 
2 N L \mu \ln N> 1, 
\end{equation} 
which is clearly satisfied in Figure \ref{fig:M10m4N15}. However, 
the usual view of clonal interference as an impediment to the simultaneous 
fixation of different beneficial mutations which slows down adaptation  
\citep{Gerrish:1998,Wilke:2004} relies on a situation in which the fitness 
effects are essentially additive, and hence strong (sign) epistasis is absent. 
In rugged fitness landscapes, on the other hand, 
the presence of several competing genotypes increases the likelihood of 
finding high 
fitness genotypes. This effect is thus seen to speed up the adaptive process  
compared to the  
regime where beneficial mutations arise and fix sequentially, which we  
consider next.

\begin{figure} 
 \centerline{\includegraphics[width=14cm,angle=270]{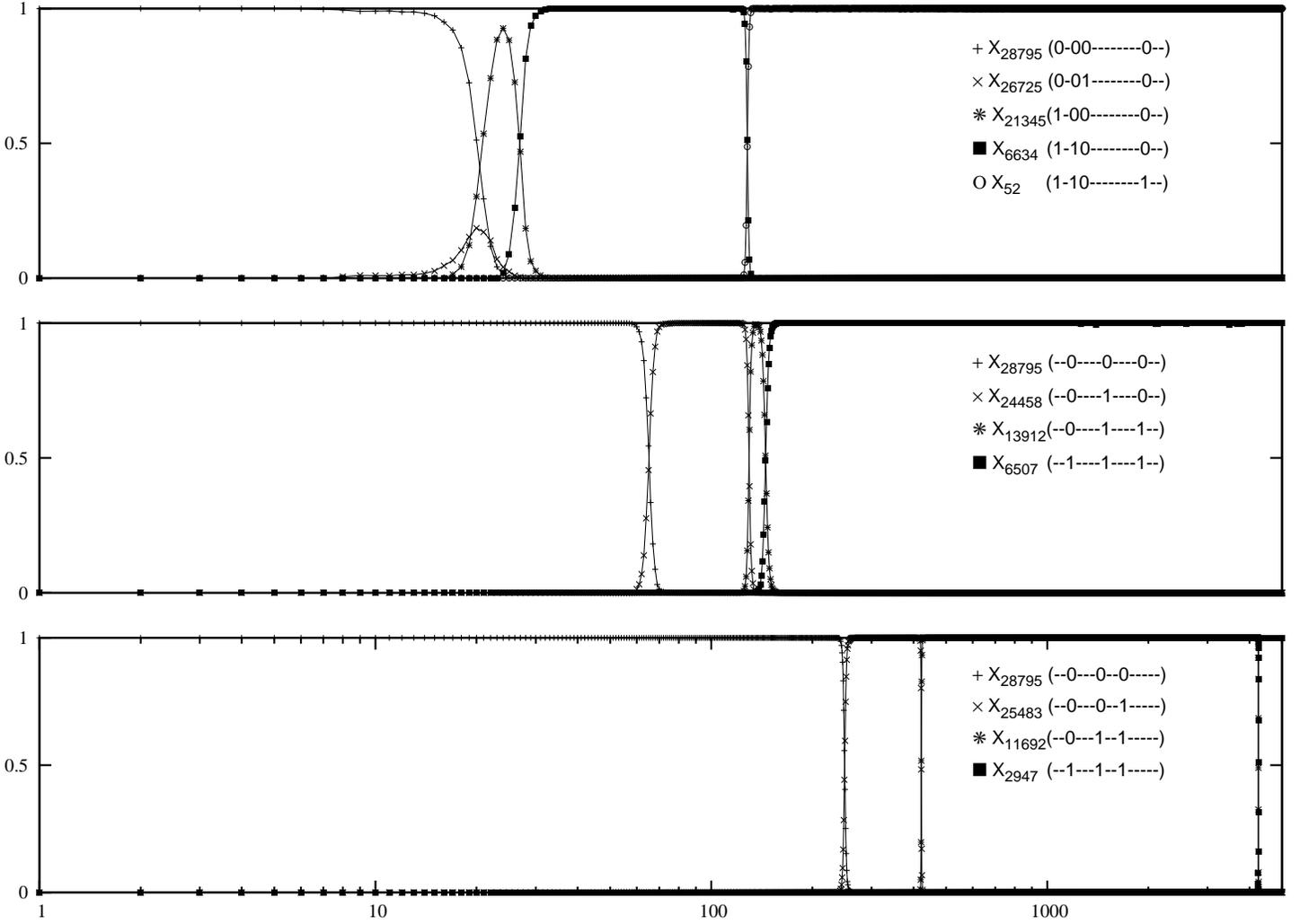}} 
\caption{\label{fig:M10N15}Population evolution when $L N \mu 
  \ll 1$ for $L=15, N=2^{10}$ with $\mu=10^{-5}$,   
  $10^{-6}$ and $10^{-7}$ (top to bottom). The mutants with 
  $X_{\mathrm{rank}}(\sigma) \geq 
  0.005$ are shown.} 
\end{figure}

{\noindent \it Adaptive walk:}  
The above discussion of course is contingent  
on the fact that several genotypes are available to explore the landscape.  
We finally consider the case in which  
the rate $L N \mu$ at which the new mutants appear is very small. Then   
the time $(L N \mu)^{-1} \gg 1$ required to produce a new mutant is very  
large, and the competing mutants are not produced enabling the population at  
the currently occupied genotype to reach a fraction unity. The population is  
thus localised at a single  
sequence at all times unlike in the previous cases where this happened  
only at long times. In 
Figure \ref{fig:M10N15}, the dynamics in the regime $L N \mu < 1$ are shown  
for three different values of $\mu$ with fixed $L$ and $N$. The  
effect of decreasing $\mu$ is similar to the quasispecies model in that  
the adaptive events are delayed and the polymorphism is reduced. Since the  
dynamics are now stochastic, the trajectories are different and an averaging  
is required to deduce the effect on fitness.  
 
At short times, the number of occupied genotypes decreases with decreasing  
mutation probability. At late times, however, the population can be  
associated with a single sequence for large $\mu$ also  
due to a reduction in $Q(W)$.  
In the topmost panel of Figure \ref{fig:M10N15}, the left hand side 
of (\ref{clonal}) is about 2, and correspondingly several genotypes coexist 
at early times.  
For $L N \mu \ll 1$, as in the bottom panel of Figure \ref{fig:M10N15},  
the population shifts as a whole by one Hamming distance. Since the mutation  
probability is small, to a first  
approximation, the population is likely to move only by one step and the  
hops to larger distances can be neglected. Thus, the population keeps moving   
one step uphill on the rugged landscape until it encounters a local  
peak whereupon this adaptive walk stops. The typical length of this walk  
is $\ln L \approx 3$ for $L=15$ \citep{Flyvbjerg:1992,Kauffman:1993}.  
For $\mu=10^{-7}$ in the bottom panel of  
Figure \ref{fig:M10N15}, the population  
reaches the sequence with rank $2947$ which is a local peak.  
The time to escape this sequence  
to a fitter one in the shell at Hamming distance two is of order  
$(L N \mu^{2})^{-1}$ (as discussed above) 
which for our choice of parameters will require   
about $10^{10}$ time steps. For small $N$ and $\mu$ that we  
consider here, it may be possible  
for a valley mutant to get fixed before the next local peak does.  
This requires that the time to fix a valley mutant is smaller than the  
time $\sim (N \mu)^{-1}$ to produce its one-mutant neighbor with  
fitness $W_{\mathrm{loc,f}}$ \citep{Carter:2002,Nowak:2004}. The valley  
mutant fixation  
time is exponentially large in $N$ if the mutant fitness  
$W_{\mathrm{mut}} \ll W_{\mathrm{loc,i}}$, while it is of order $N$  
for the near neutral case. Clearly, the above  
requirement can be met only when the population escapes  
through a ``shallow'' valley which is a rather unlikely scenario in a rugged  
landscape.   
 
Before the population gets trapped at a local peak, the   
dynamics can be described by the mutational landscape model  
\citep{Gillespie:1984} which applies to   
a genetically homogeneous population undergoing beneficial mutation with  
a very low probability.  
As pointed out in \citet{Orr:2002}, the behavior of the population  
undergoing an adaptive walk is neither deterministic  
 nor completely random in  
that each (better) mutant would be equally likely to get  
fixed. In fact, each one-mutant neighbor better than  
the currently occupied one has a probability to get fixed given by  
\begin{equation} 
P_{\mathrm{fix}}(\sigma \vert \sigma^{(0)})=\frac{\Pi (\sigma \vert 
\sigma^{(0)})}{\sum_{\sigma^\prime} \Pi (\sigma^\prime \vert \sigma^{(0)})} , 
\end{equation} 
where the sum is over the fitter nearest neighbors of $\sigma^{(0)}$, 
and the unnormalized fixation probability is given by  
\begin{equation} 
\label{fixation} 
\Pi (\sigma \vert \sigma^{(0)})= \frac{s(\sigma,\sigma^{(0)})}{1+s(\sigma,\sigma^{(0)})}=1-\frac{W(\sigma^{(0)})}{W(\sigma)} 
\end{equation} 
for large $N$ \citep{Durrett:2002}.  
In the last panel of Figure \ref{fig:M10N15}, the probability for the  
sequence ranked $25483$ to get fixed is $\approx 0.049$ which is almost  
half of the fixation probability $\approx 0.095$ of the  
best available sequence with rank $4688$.

\bigskip 
\centerline{DISCUSSION} 
\bigskip 
 
In this article, we posed the question under what conditions biological 
evolution is predictable. To answer this, we studied the dynamics  
of a finite population $N$ within a mutation-selection  
model defined  
on the space of binary genotype sequences of length $L$.  
This work thus considers $L$ loci models, unlike that of \citet{Rouzine:2001}  
which focuses on the one locus problem.  
Our simulations also differ from those of \citet{Wahl:2000} where the dynamics  
are described by the quasispecies equation (\ref{quasi})  
as long as the population fraction $X(\sigma,t)$  
exceeds $1/N$ and if the fraction falls below this cutoff, an  
individual is added to sequence $\sigma$ with a certain probability. We have 
instead simulated the full 
stochastic process defined by Wright-Fisher  
dynamics  which allows us to track the exact evolutionary path  
of any mutant. The fitness  
landscape under consideration is highly epistatic with many local optima.

We classified the various evolutionary regimes using a parameter  
$d_{\mathrm{eff}}$ defined in (\ref{deff}) which has been obtained under  
the assumption of strong selection. Usually the boundary  
between deterministic and stochastic evolution is defined by  
the product $N \mu$ \citep{Johnson:1995,Wahl:2000,Rouzine:2001};  
as most of these theories are based on one-locus  
models \citep{Johnson:1995,Rouzine:2001}, the description 
in terms of $N \mu$  
suffices. We are instead dealing with the whole sequence space in which  
mutations can occur to a distance greater than unity depending on the  
population size $N$ and mutation probability $\mu$. This  
requires a description in terms of  
the distance $d_{\mathrm{eff}}$ which measures the typical distance to  
which the mutants can spread.  
The boundary $N \mu=1$ is included  
in our description as this corresponds to $d_{\mathrm{eff}}=1$.  
However, in contrast to the product $N \mu$, the logarithmic dependence 
of (\ref{deff}) implies that moderate changes 
in $d_\mathrm{eff}$ require enormous changes of $N$ or $\mu$.  
 
Our conclusions summarised in Table \ref{regime} fall into three broad  
categories. The infinite population 
case with  $d_{\mathrm{eff}}=L$ is described by the deterministic quasispecies  
model \citep{Eigen:1971}.  
Given the fitness landscape and the starting point, one can predict the 
 path taken by 
the initially unfit population to a peak in the landscape. For finite  
populations with $ d_{\mathrm{eff}} \gtrsim 1$, although the long time course  
is determined by  
stochastically occurring rare mutations, it is possible to predict the  
trajectory until a time $T_{\times}$ (Equation \ref{Ttimesf}) 
that increases with $L$ and $N$ using the  
deterministic prescription \emph{locally}.  
We emphasize that  
the dynamics described by the local quasispecies theory which applies to  
shells of size $d_{\mathrm{eff}}$ centred about the \emph{current} $\sigma^*$  
is different from the quasispecies theory applied to the Hamming space  
restricted to the shells up to the one in which the local  
peak is located. This is simply because the initial population  
$\mu^d$ at the local peak in question can be smaller than $1/N$.  
The intuitive picture provided by the local  
quasispecies theory is in fact equivalent to the description in  
\citet{Wahl:2000} where quasispecies is applied to full  
space provided the lower cutoff $1/N$ is imposed. The 
viewpoint that quasispecies dynamics can be useful in understanding 
the behavior of finite populations has been expressed by other  
authors also (see, for example \citet{Wilke:2005}).  
 
The local quasispecies description breaks down when the population  
fails to find a genotype better than the currently occupied one within  
distance $d_{\mathrm{eff}}$. Then rare mutations (of  
the order $\mu^{d_{\mathrm{eff}}+1}$) that allow  
the population to access a distance $> d_{\mathrm{eff}}$  
play an important role. On rugged landscapes, the population can escape this  
situation either by double mutations (for $N \mu \sim 1$) or tunneling  
through the low fitness mutants \citep{Iwasa:2004,Weinreich:2005}.  
Large populations are able to cross a fitness valley much more rapidly than  
expected on the basis of the adaptive walk picture, in which the fixation of  
a deleterious mutation is exponentially unlikely  
\citep{Nimwegen:2000,Gavrilets:2004,Weinreich:2005}. The reason is that in a  
large population the less fit  
genotypes connecting the two fitness peaks are always  
present in some number, enabling the population to climb the new peak  
without ever in its entirety residing in the valley. This is similar to the  
peak shift mechanism found in the quasispecies model, where 
all possible mutants are alway present in the population  
\citep{Jain:2005,Jain:2006}.  
 
To summarise, there is a crossover in the dynamics when  
$d_{\mathrm{eff}} \gtrsim 1$ from a deterministic quasispecies type  
dynamics to stochastic dynamics in which stochastic escapes occur.  
For RNA virus with typical  
population size $N \sim 10^{6}$ and mutation  
probability $\mu \sim 10^{-3}$ per base per generation in a  
genome of about thousand bases   
\citep{Lazaro:2006}, these parameters give  
$d_{\mathrm {eff}} \approx 2$ which suggests that the local quasispecies 
dynamics operate in the finite viral populations for short 
times. This scenario is expected to hold good for HIV also for which the  
product $N \mu \sim 1$ \citep{Rouzine:1999}.  
 
For $N \mu < 1$, the dynamics are stochastic right from the start.  
The long time dynamics are expected to be qualitatively similar to  
that discussed above. But the short time dynamics differ considerably  
and depend on the number of one-mutant neighbors.  
While many analytical results are available for the adaptive walk limit  
\citep{Gillespie:1984,Kauffman:1993}, the parameter regime  
when $N L \mu$ is not too small on epistatic landscapes requires  
further attention. In experiments on   
\emph {E. Coli} which has $L \sim 10^{6}$, $\mu \sim 10^{-10}$ and  typical  
colony sizes of order $10^6$, 
$N \mu \ll 1$ but $L N \mu \gg 1$, which hints at the stochastic nature of  
the bacterial evolution. This behavior has been seen in the experiments by  
the Lenski group in which the fitness 
of bacterial populations evolving under identical conditions diverged 
in time \citep{Korona:1994,Lenski:1994}.

In this article, we have provided a unified picture of the nature of the  
evolutionary process.  
As our models are defined on sequence space, this constitutes a step  
towards realistic modeling of the biological evolution occurring in the  
genotypic space. Inclusion of other relevant factors such as recombination  
could be the next step in our understanding of genetic evolution.

\begin{table} 
\begin{center} 
\begin{tabular}{|c|c|c|c|} 
\hline 
& $d_{\mathrm{eff}}=L$ &  $1 \leq d_{\mathrm{eff}} < L$ & $d_{\mathrm{eff}} < 1$ \\ 
\hline 
Behavior   & Deterministic & Crossover deterministic $\to$ stochastic  & Stochastic \\ 
\hline 
Regime  & Quasispecies & $t < T_{\times}$: Local quasispecies & $L N \mu \gtrsim 1$:  
Clonal interference \\ 
& & $t > T_{\times}$: Valley crossing or hopping &   $L N \mu < 1$: Adaptive walk\\ 
\hline 
\end{tabular} 
\end{center} 
\caption{Summary of regimes in evolution on rugged landscapes where  
$d_{\mathrm{eff}}=\ln N/\vert \ln \mu \vert$.} 
\label{regime} 
\end{table}

Acknowledgement: This work has been supported by Deutsche Forschungsgemeinschaft  
within SFB/TR 12 \textit{Symmetries and Universality in Mesoscopic Systems}  
and SFB 680 \textit{Molecular Basis of Evolutionary Innovations.} 
KJ acknowledges financial support from the Israel Science Foundation and thanks  
ITP, University of K\"oln for kind hospitality during a visit.  
JK is grateful to U. Gerland for useful discussions.  
The authors would also like to thank the Isaac Newton Institute for Mathematical Sciences where a part of this work was carried out, and are grateful to an anonymous referee for valuable comments which were helpful in sharpening the definition of distance $d_{\rm{eff}}$. 
\bigskip 
 

\end{document}